\documentclass[aps,prb,twocolumn,superscriptaddress]{revtex4-2}
\usepackage{graphicx}
\usepackage{color}
\usepackage{hyperref}
\usepackage{dsfont}
\usepackage[]{changes}
\definechangesauthor[name={parkman}, color=red]{XuTao}

\begin{document}


\title{Type-II antiferromagnetic ordering in double perovskite oxide Sr$_2$NiWO$_6$}


\author{Cheng Su}
\thanks{These authors contributed equally to this work.}
\affiliation{School of Physics, Beihang University, Beijing 100191, China}

\author{Xu-Tao Zeng}
\thanks{These authors contributed equally to this work.}
\affiliation{School of Physics, Beihang University, Beijing 100191, China}

\author{Kaitong Sun}
\thanks{These authors contributed equally to this work.}
\affiliation{Joint Key Laboratory of the Ministry of Education, Institute of Applied Physics and Materials Engineering, University of Macau, Avenida da Universidade, Taipa, Macao SAR 999078, China}

\author{Denis Sheptyakov}
\affiliation{Laboratory for Neutron Scattering and Imaging, Paul Scherrer Institut, CH-5232 Villigen-PSI, Switzerland}

\author{Ziyu Chen}
\affiliation{School of Physics, Beihang University, Beijing 100191, China}

\author{Xian-Lei Sheng}
\affiliation{School of Physics, Beihang University, Beijing 100191, China}

\author{Haifeng Li}
\email{haifengli@um.edu.mo}
\affiliation{Joint Key Laboratory of the Ministry of Education, Institute of Applied Physics and Materials Engineering, University of Macau, Avenida da Universidade, Taipa, Macao SAR 999078, China}

\author{Wentao Jin}
\email{wtjin@buaa.edu.cn}
\affiliation{School of Physics, Beihang University, Beijing 100191, China}


\date{\today}

\begin{abstract}
Magnetic double perovskite compounds provide a fertile playground to explore interesting electronic and magnetic properties. By complementary macroscopic characterizations, neutron powder diffraction measurements and first-principles calculations, we have performed comprehensive studies on the magnetic ordering in the double perovskite compound Sr$_2$NiWO$_6$. It is found by neutron diffraction to order magnetically in a collinear type-II antiferromagnetic structure in a tetragonal lattice with $k$ = (0.5, 0, 0.5) below $T\rm_N$ = 56 K. In the ground state, the ordered moment of the spin-1 Ni$^{2+}$ ions is determined to be 1.9(2) $\mu\rm_{B}$, indicating a significant quenching of the orbital moment. The Ni$^{2+}$ moments in Sr$_2$NiWO$_6$ are revealed to cant off the $c$ axis by 29.2$^{\circ}$, which is well supported by the first-principles magnetic anisotropy energy calculations. Furthermore, the in-plane and out-of-plane next-nearest-neighbor superexchange couplings ($J\rm_2$ and $J\rm_{2c}$) are found to play a dominant role in the spin Hamiltonian of Sr$_2$NiWO$_6$, which accounts for the stabilization of the type-II AFM structure as its magnetic ground state. 
\end{abstract}


\maketitle

\section{Introduction}
Perovskite oxide $AB$O$_3$ has drawn great attention and interest during the past few decades \cite{chakhmouradian2014,attfield2015}, owing to their highly tunable physical properties and promising applications in electronic or spintronic devices \cite{fakharuddin2022perovskite,privitera2021perspectives}, fuel cells \cite{huang2006double}, solar cells \cite{green2014emergence} and so on, 
arising from their highly flexible chemical and structural properties. As a variant of the perovskite structure, the $B$-site ordered double-perovskite (DP) oxides with the general formula of $A_2BB'$O$_6$ ($A$ being a divalent or trivalent metal, $B$ and $B'$ being transition-metal ions alternately arranged in a rock-salt structure and surrounded by corner-sharing oxygen octahedra) have been the focus of intensive studies in recent years \cite{vasala2015a2b,hossain2018overview}, because of various intriguing electronic and magnetic properties that may be realized in this large family. The most well-known example is Sr$_2$FeMoO$_6$ with a high Curie temperature of $T\rm_{C}$ $\sim$ 420 K and a large room-temperature magnetoresistance, as a result of the strong 3$d$-4$d$ hybridization between the Fe$^{3+}$ and Mo$^{5+}$ ions \cite{kobayashi1998room,tomioka2000magnetic}. Such a hybridization between the $B$ and $B'$ ions is also evidenced in 3$d$-5$d$ DP rhenates and iridates \cite{kobayashi1999intergrain,kolchinskaya2012magnetism,lee2018hybridized,jin2022magnetic}, in which the strong spin-orbit coupling on 5$d$ ions may give rise to anisotropic and bond-dependent magnetic interactions. 

   For magnetic $B$-site ions located in the center of the $B$O$_6$ octahedra, although the direct exchange interaction is negligible due to the large distance between them, the nearest-neighbor (NN) and next-nearest-neighbor (NNN) superexchange interactions can take place over a 90$^{\circ}$ $B$-O-($B'$)-O-$B$ path and a 180$^{\circ}$ $B$-O-$B'$-O-$B$ path, respectively. Typically, the $B$-site ordered DP with a single magnetic sublattice shows a low-temperature antiferromagnetic (AFM) ordering. Depending on the identities of $B$ and $B'$ ions and the relative strength of the NN ($J_1$) and NNN ($J_2$) interactions, different types of AFM structures have been observed \cite{battle1989crystal,rodriguez2002crystal,makowski2009coupled}. As two distinct examples, the magnetic ground state of Sr$_2$CuTeO$_6$ is a N\'{e}el-type AFM ordering as expected for $J_2$ \textless $J_1$ \cite{koga2016magnetic}, while Sr$_2$CuWO$_6$ exhibits a type-II AFM ordering as expected for $J_2$ \textgreater $J_1$ \cite{vasala2014magnetic}. Accordingly, they show a quasi-two-dimensional (2D) and three-dimensional (3D) nature, respectively, associated with the magnetism of the $S$ = 1/2 Cu$^{2+}$ ions.

	Sr$_2$NiWO$_6$ (SNWO) is a $B$-site ordered DP oxide, whose detailed magnetic structure of the $S$ = 1 Ni$^{2+}$ moments is unknown so far. The divalent magnetic Ni$^{2+}$ ions and hexavalent diamagnetic W$^{6+}$ ions occupy the $B$ and $B'$ sites,   respectively, forming interpenetrating face-centered cubic (FCC) lattices. It crystallizes in a tetragonal structure (space group $I$4/$m$) at the ambient condition \cite{nomura1966magnetic,IWANAGA2000SNWOandSNTO,BLUM2015Fluxgrowth}, and undergoes a tetragonal-to-cubic structural phase transition above 300 $^{\circ}$C \cite{gateshki2003XRDforPT,zhou2005structural}. Interestingly, SNWO was found to order antiferromagnetically at a rather high N\'{e}el temperature of $T\rm_{N}$ = 54-59 K \cite{nomura1966magnetic,IWANAGA2000SNWOandSNTO,Todate99,BLUM2015Fluxgrowth}, despite a very long superexchange path (Ni-O-W-O-Ni) of $\sim$ 8 \AA. Although an $ab$ $initio$ calculation predicts the type-II AFM ordering (AFM-II structure) as its magnetic ground state \cite{rezaei2019ab}, which is consistent with the expectation from the inelastic neutron scattering measurements \cite{Todate99}, and on the other hand a similar DP oxide Ba$_2$NiWO$_6$ was proved to be of a type-II AFM structure as well \cite{cox1967neutron}, a direct confirmation using magnetic neutron diffraction and detailed knowledge of the moment size and direction in SNWO are still lacking, which impedes a thorough understanding about the spin interactions and the origin of AFM ordering in SNWO. 

	In this work, we have conducted comprehensive studies on the magnetic ordering in SNWO, combining macroscopic and microscopic experimental probes and first-principles calculations. Using neutron diffraction as the microscopic probe, we have provided a solid evidence that SNWO orders magnetically in a type-II AFM structure with $k$ = (0.5, 0, 0.5) below $T\rm_N$ = 56 K, in good agreement with the previous theoretical prediction.  At the base temperature, the Ni$^{2+}$ moments in SNWO are found to cant off the $c$ axis by 29.2$^{\circ}$, well supported by the first-principles magnetic anisotropy energy calculations. The strengths of the superexchange couplings are also estimated to understand the origin of the experimentally determined type-II AFM structure. 

\section{Methods}
	Polycrystalline samples of Sr$_2$NiWO$_6$ were synthesized by a standard solid-state reaction method as reported in Ref. \cite{BLUM2015Fluxgrowth}. NiWO$_4$, the precursor material, was firstly synthesized by sintering a stoichiometric mixture of WO$_3$ (99.99\%) and NiO (99.9\%) in air, at 1000 $^{\circ}$C for 10 h. Then the synthesized NiWO$_4$ powder was mixed with SrCO$_3$ (99.95\%) with a stoichiometric ratio of 1:2, and sintered in air at 1400 $^\circ$C for 8 h. The powder mixture in each procedure mentioned above needs to be thoroughly ground and pelletized before the sintering. The phase purity was checked by a room-temperature x-ray diffraction (XRD) measurement on a Bruker D8 ADVANCE diffractometer with Cu-K$\alpha$ radiation ($\lambda$ = 1.5406 \AA). 

	The neutron powder diffraction (NPD) experiments were carried out on the High Resolution Powder diffractometer for Thermal neutrons (HRPT) \cite{HRPT} at the Swiss Neutron Spallation Source (SINQ), at the Paul Scherrer Institute in Villigen, Switzerland. Powder sample of SNWO with a total mass of 5.5 grams was loaded into a 8 mm-diameter vanadium can. The diffraction patterns were collected using wavelengths $\lambda$ = 1.1545\AA, 1.494\AA \ and 2.45\AA \,  in the temperature range from 1.5 to 300 K. The shorter wavelengths cover a wider $Q$-range suitable for refining the structural parameters, while the longer wavelength provides a higher resolution in 
	the lower $Q$-range for a more reliable determination of magnetic structure. Refinements of the nuclear and magnetic structures were conducted using the FULLPROF program suite \cite{Rodriguez-Carvajal_93}.

	First-principles calculations were performed on the basis of density-functional theory (DFT) using the generalized gradient approximation (GGA) in the form proposed by Perdew $\mathit{et}$ $\mathit{al}$. \cite{perdew1996}, as implemented in the Vienna $ab$ $initio$ Simulation Package (VASP) \cite{kresse1996,joubert1999}. The energy cutoff of the plane wave was set to 500 eV. The energy convergence criterion in the self-consistent calculations was set to \textcolor{black}{10$^{-6}$} eV. A $\Gamma$-centered Monkhort-Pack $k$-point mesh with a resolution of 2$\pi$\texttimes{}0.03\textcolor{black}{{} \AA{}$^{-1}$} was used for the first Brillouin zone sampling. To account for the correlation effects for Ni, we adopted the GGA + $\mathit{U}$ method \cite{anisimov1991} with the value of $\mathit{U}$ = 5 eV, which is commonly used in studying nickel compounds \cite{Bengone2000}.

\section{Results and Discussions}
	
     \begin{figure*}[htb]
		\centering
		\includegraphics[width=\linewidth]{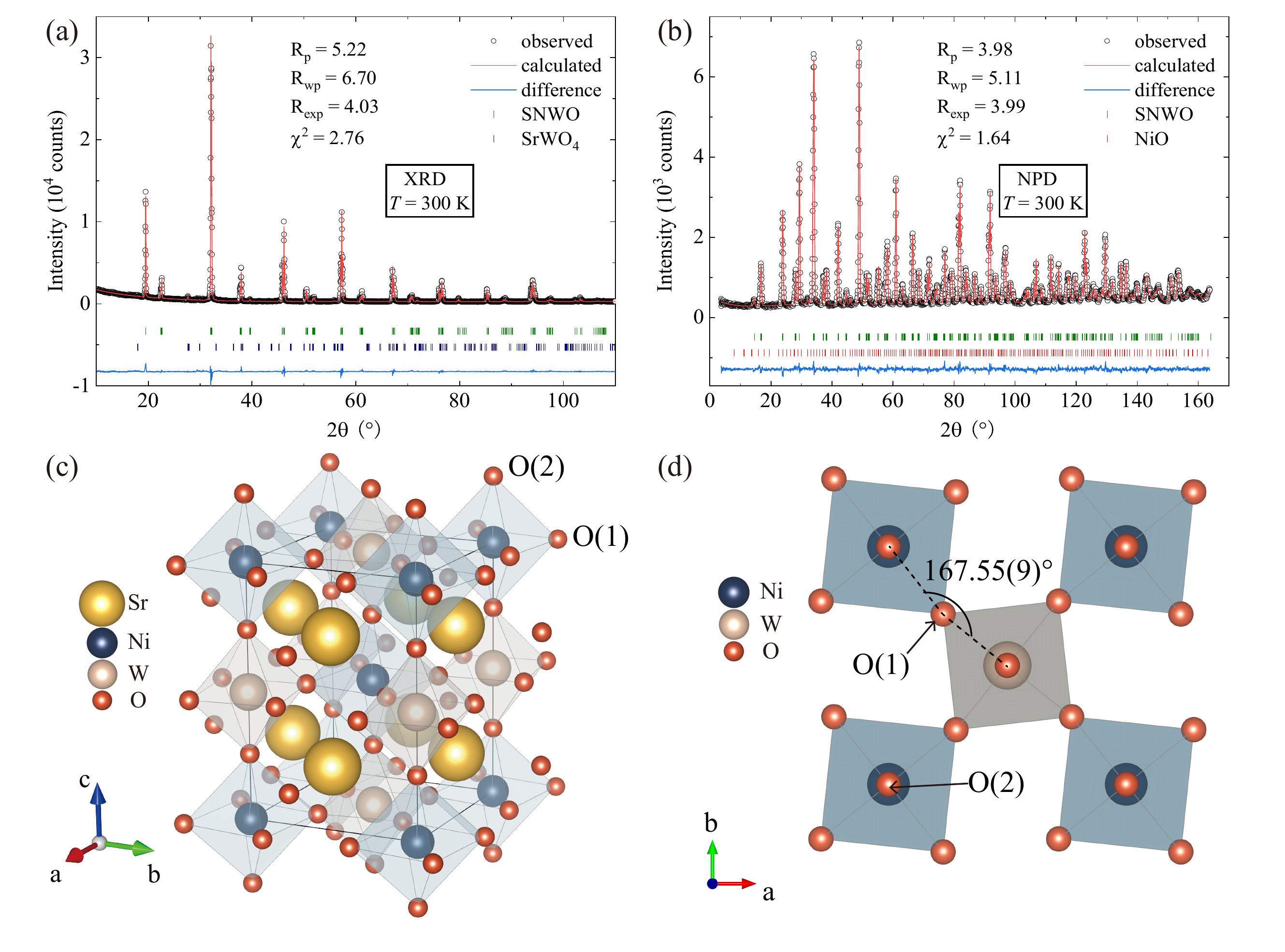}
		\caption{\label{fig:XRD} Room-temperature XRD (a) and NPD (b) patterns of the polycrystalline SNWO sample and the corresponding crystal structure (c, d). In (a, b), the black circles represent the observed intensities and the red solid line is the calculated pattern according to the Rietveld refinement. The difference between the observed and calculated intensities is shown as the blue line at the bottom. The olive, red and navy vertical bars indicate the Bragg reflections from the SNWO main phase, NiO and SrWO$_4$ impurities, respectively. (d) illustrates the projection of the lattice of SNWO onto the $ab$ plane, showing the out-of-phase rotations of NiO$_6$ and WO$_6$ octahedra along the $c$ axis and the resultant in-plane Ni-O(1)-W bond angle of 167.55$^{\circ}$ at 300 K.}
	\end{figure*}

	\subsection{Structural characterizations}
	
	\begin{table}
	\caption{\label{Table:atom info} Refinement results of the lattice parameters, atomic coordinates and thermal factors of SNWO at 300 K and 1.5 K, respectively. (Space group $I4/m$, $Z$=2)}
	
	\begin{ruledtabular}
		\resizebox{\columnwidth }{!}{
		\begin{tabular}{ccccc}
			$T$ = 300 K & \multicolumn{4}{c}{ $a=b=5.5598(1)$\AA, $c=7.9173(2)$\AA  }\\
			Atom (site) & $x$      & $y$         & $z$    & $B\rm_{iso}({\AA}^2)$\\
			\hline
			O(1) (8h) & 0.2843(3) & 0.2298(3) & 0         & 0.79(2)\\
			O(2) (4e) & 0         & 0         & 0.2566(4) & 0.87(2)\\
			W (2b)  & 0         & 0         & 0.5       & 0.57(4)\\
			Sr (4d) & 0         & 0.5       & 0.25      & 0.60(1)\\
			Ni (2a) & 0         & 0         & 0         & 0.20(2)\\
			\hline
			\\
			$T$ = 1.5 K & \multicolumn{4}{c}{ $a=b=5.5409(2)$\AA, $c=7.9264(2)$\AA }\\
			Atom (site) & $x$      & $y$         & $z$    & $B\rm_{iso}({\AA}^2)$\\
			\hline
			O(1) (8h) & 0.2912(3) & 0.2246(3) & 0         & 0.38(2)\\
			O(2) (4e) & 0         & 0         & 0.2576(4) & 0.33(3)\\
			W (2b)  & 0         & 0         & 0.5       & 0.42(6)\\
			Sr (4d) & 0         & 0.5       & 0.25      & 0.17(2)\\
			Ni (2a) & 0         & 0         & 0         & 0.05(3)\\
			\end{tabular}
		}
	\end{ruledtabular}
	
	\end{table}
	
	Figure 1(a) and 1(b) show the room-temperature XRD and NPD patterns of the synthesized polycrystalline SNWO sample, respectively, which can be well fitted using the reported tetragonal phase of SNWO \cite{IWANAGA2000SNWOandSNTO,BLUM2015Fluxgrowth,gateshki2003XRDforPT} with satisfactory $R$ factors. Tiny amounts of impurity phases were detected in x-ray (SrWO$_4$, 0.9\% wt) and neutron (NiO, 0.5\% wt) powder diffraction data and included into the corresponding Rietveld refinements. The contribution of SrWO$_4$ to the neutron powder data was however so weak that it could be omitted in the refinements.

   Due to the relatively large coherent neutron scattering length of oxygen atoms, the room-temperature structural parameters of SNWO including all atomic coordinates can be precisely determined by Rietveld refinements to the NPD pattern shown in Fig. 1(b) and listed in Table \ref{Table:atom info}. According to our refinements, the atomic coordinates are consistent with those reported \cite{IWANAGA2000SNWOandSNTO,gateshki2003XRDforPT,XU2017Solgel} and no obvious site mixing between Ni and W was observed. Thus, the actual structure of the synthesized SNWO is indeed a rock-salt type $B$-site ordered DP (see Fig. 1(c)). 

   As illustrated in Fig.1(d), the adjacent NiO$_6$ and WO$_6$ octahedra display out-of-phase rotations around the $c$ axis (denoted as an $a^0a^0c^-$ rotation pattern in the well adopted Glazer notation in discribing the perovskite structures \cite{Glazer72}), yielding an in-plane Ni-O(1)-W bond angle of 167.55$^{\circ}$ and a tetragonal symmetry at 300 K, in contrast to a non-distorted 180$^{\circ}$ Ni-O(1)-W bond angle associated with the cubic symmetry at very high temperature \cite{gateshki2003XRDforPT,zhou2005structural}. Such a lattice distortion has been widely reported in various DP compounds \cite{vasala2015a2b}, which releases the stress caused by the mismatch of ionic radius on the $B$ and $B'$ sites and thus provides a great tolerance of the DP structure to accommodate most elements in the periodic table. 
	
	\subsection{Macroscopic magnetic properties}
	
	\begin{figure}
			\centering
			\includegraphics[width=8.6cm]{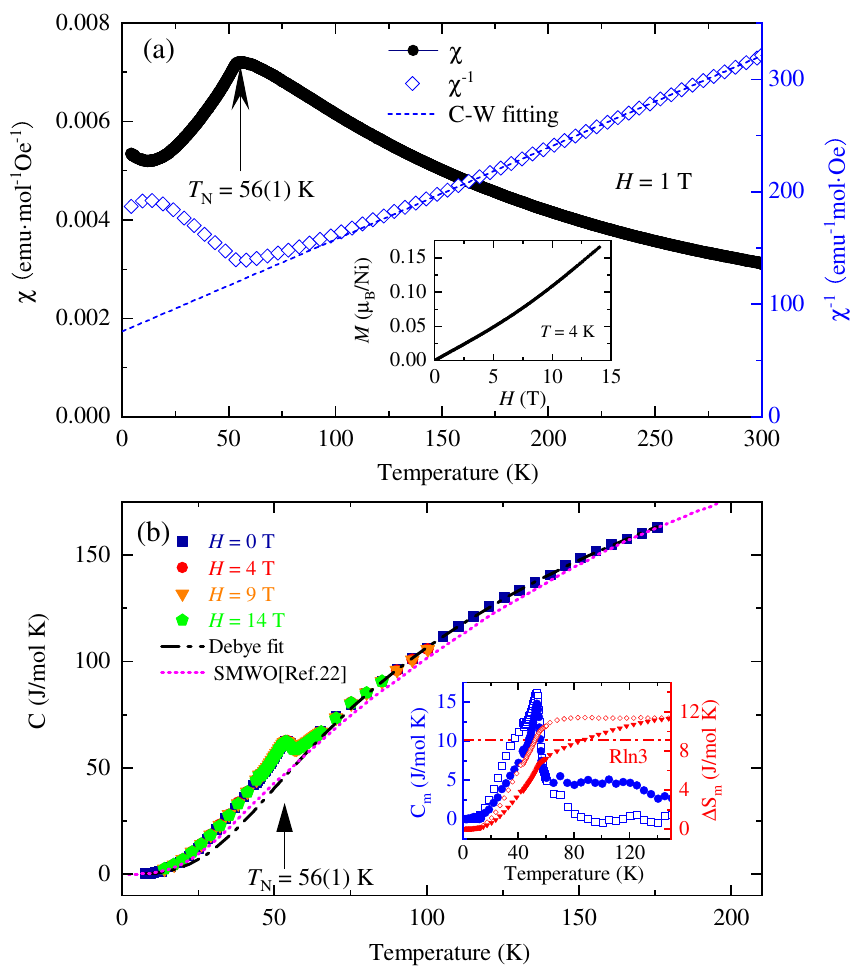}
			\caption{\label{fig:PhysProp}  
									(a) DC magnetic susceptibility ($\chi$, black circles) and inverse susceptibility ($1/\chi$, blue squares) of the polycrystalline SNWO sample, measured in a magnetic field of 1 T.  The dashed line represents the Curie-Weiss fitting to $1/\chi$ from 200 to 300 K. The inset shows the isothermal magnetization curve measured at 4 K. 
									(b) Molar specific heat $C$ of SNWO measured in the magnetic field of 0, 4, 9, and 14 T, respectively. The dashed line represents a fitting to the phonon contribution $C\rm_{ph}$ to the zero-field specific heat. The black dashed line represents the two-component Debye model fitting and the magenta dotted line represents the specific heat of polycrystalline $\rm{Sr_2MgWO_6}$ extracted from Ref. \cite{BLUM2015Fluxgrowth}. The inset shows the magnetic specific heat $C\rm_{m}$ (blue squares from the former approach and circles from the latter) after the phonon subtraction and the change of magnetic entropy $\Delta S\rm_{m}$ (red diamonds from the former approach and triangles from the latter) calculated by integrating $C\rm_{m}$/$T$. For comparison, the expected $\Delta S\rm_{m}$ value of $R$ln3 is also marked.
				}
			\end{figure}

	The temperature dependence of the dc magnetic susceptibility of the polycrystalline SNWO measured in an applied magnetic field of 1 T is shown in Fig. 2(a), which clearly indicates a typical AFM transition at $T\rm_{N}$ = 56(1) K, consistent with the previously reported N\'{e}el temperature of 54-59 K \cite{nomura1966magnetic,IWANAGA2000SNWOandSNTO,Todate99,BLUM2015Fluxgrowth}. By performing a Curie-Weiss fitting to the inverse magnetic susceptibility (1/$\chi$) in the paramagnetic state from 200 to 300 K, an effective magnetic moment of $\mu\rm_{eff}$ = 3.127(2) $\mu\rm_{B}$ for the Ni$^{2+}$ ions and a Curie-Weiss temperature of $\theta$ = $-$92.8(3) K are obtained. The $\mu\rm_{eff}$ value is close to the spin-only moment value of 2.83 $\mu\rm_{B}$ for the spin-1 Ni$^{2+}$ ions, suggesting the significant quenching of their orbital angular momentum in the octahedral coordination formed by surrounding oxygen ions. 

	The large negative value of $\theta$ close to $-$10$^2$ K suggests a very strong dominant AFM interaction between the Ni$^{2+}$ moments in SNWO, which is further supported by the isothermal magnetization curve measured at 4 K, as shown in the inset of Fig. 2(a). Up to 14 T, the measured magnetization is still far below the saturated value of $\mu\rm_{S}$ = 2 $\mu\rm_{B}$ per Ni$^{2+}$ ion with the spin-only moment (Land\'{e} factor $g$ = 2), as a free spin-1 ion coupling with a 14 T magnetic field is estimated to obtain an Zeeman energy of $\sim$ 20 K, which is far below the characteristic temperature of $\theta$ = -92.8(3) K.

	In addition, as shown in Fig. 2(b), the molar specific heat ($C$) of SNWO also shows a clear anomaly at $\sim$ 54 K, supporting the long-range nature of the AFM ordering revealed in Fig. 2(a). It's worth noting that such an anomaly around $T_N$ is hardly affected by the external field up to 14 T, further corroborating the robustness of the AFM interactions in SNWO against the external perturbation.  
	To further obtain the magnetic specific heat of SNWO, two different approaches are adopted to estimate the phonon contribution.
		
	The first method is to approximate the phonon specific heat by the Debye model. A two-component model assuming two different Debye temperatures, 
		$$ C\rm_{ph} = 9\it{R}\sum_{i=\rm1}^{\rm2} {c_i (\frac{T}{\theta_{Di}})^{\rm3} \int_{\rm0}^{\theta_{Di}/T}{ \mathrm{d}x  \frac{x^{\rm4} e^x}{(e^x-\rm1)^{2\rm}} } ,}$$
	 is used with the constrain of $c_1 + c_2 = 10$, as there are 10 atoms in total in one chemical formula unit. 
	The best fit converges to the value of $c_1$ = 5.607, $c_2$ = 4.393, $\theta_{D1}$ = 293.5 K and $\theta_{D2}$ = 868.5 K. 
	Considering the large atomic mass difference in SNWO (for example W and O atoms), this two-component model can be interpreted as the division of atoms into the light and heavy groups, as widely adopted in many other materials \cite{somesh2021,ranjith2017,kini2006}. With this estimated phonon specific heat subtracted, the magnetic specific heat ($C\rm_{m}$) is obtained and the corresponding magnetic entropy change ($\Delta S\rm_{m}$) is calculated by integrating $C\rm_{m}$/$T$ over the temperature, as shown by the open squares and diamonds in the inset of Fig.2(b), respectively.
		
	The other method to subtract the phonon specific heat is to use a non-magnetic analog. Here we refer to the experimental data of polycrystalline $\rm{Sr_2MgWO_6}$ (SMWO) from Ref. \cite{BLUM2015Fluxgrowth}. By multiplying a factor of 1.106 to the specific heat of SMWO to have its high-temperature behavior overlapped with that of our SNWO, the magnetic specific heat of SNWO can be obtained as well after the subtraction and $\Delta S\rm_{m}$ is calculated accordingly, as shown by the solid circles and triangles in the inset of Fig. 2(b).
	
	Using either of the two approaches presented above, the experimental maximal change of magnetic entropy $\Delta S\rm_{m}$ is estimated to be $\sim$ 11.5 JK$^{-1}$mol$^{-1}$ up to 150 K. 
	Considering the approximation in the Debye model of taking phonon density of states as quadratic forms and the substantial difference between the atomic mass of Mg and Ni in SMWO and SNWO, both approaches cannot capture the phonon behaviors precisely. Therefore, the difference between the measured magnetic entropy change and the theoretically expected value for a spin-1 system ($\Delta S\rm_{m}$ = $R$ln(2$S$+1) = $R$ln3 = 9.134 JK$^{-1}$mol$^{-1}$) can be due to an imprecise estimation of the phonon contributions. However, the characterization about the specific heat still clearly suggests that the magnetism of Ni$^{2+}$ spins can be well described by $S$ = 1.

	\subsection{Magnetic structure determination}
	
	\begin{figure*}
		\centering
		\includegraphics[width=\linewidth]{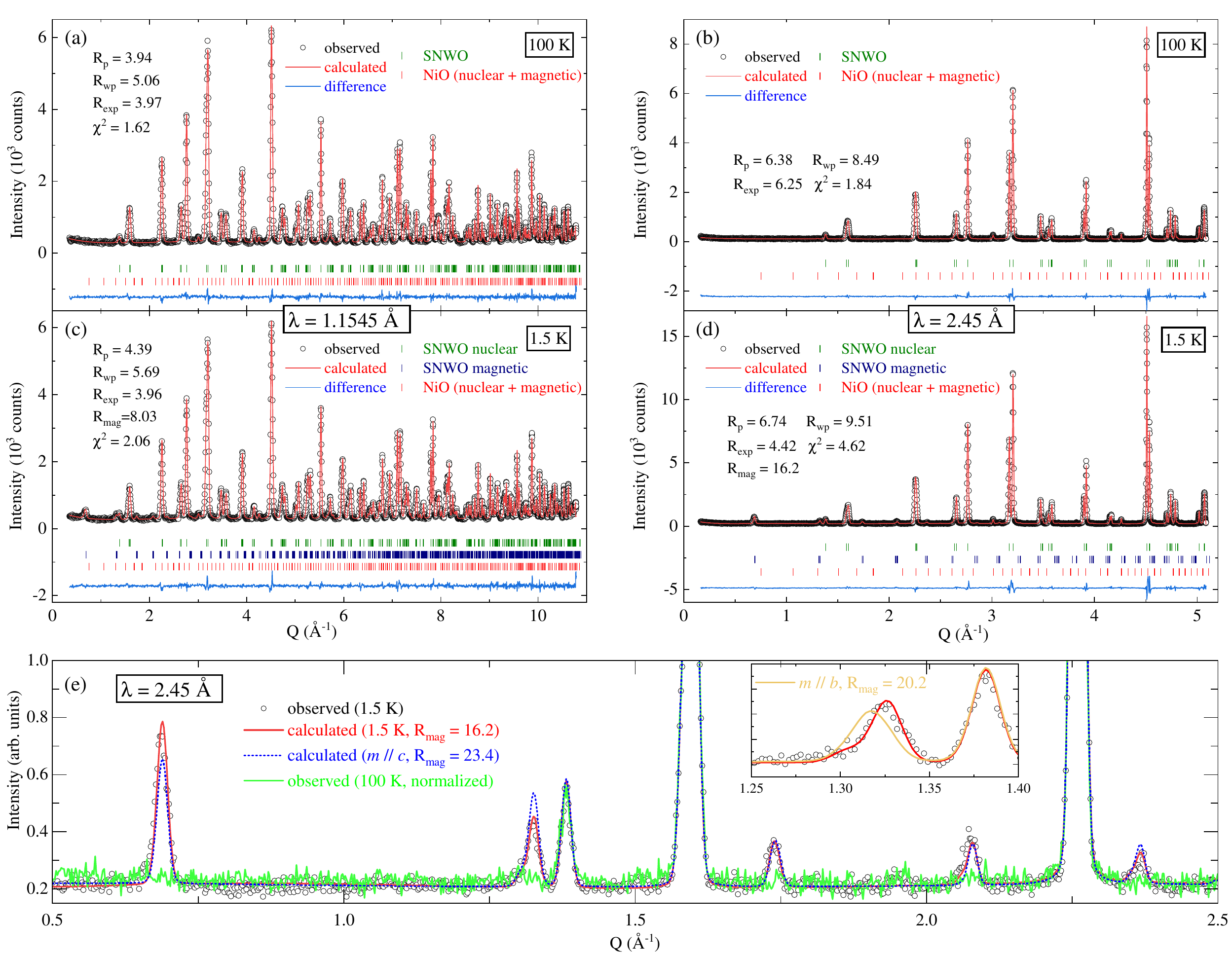}
		\caption{\label{fig:NPD} 
			NPD patterns of SNWO at 100 K (a, b) and 1.5 K (c, d) and the corresponding Rietveld refinements. The left (a, c) and right (b, d) panels show the data collected with the neutron wavelength of 1.1545 \AA$ $ and 2.45 \AA$ $, respectively. The Rietveld refinements at 1.5 K were performed adopting the magnetic structure described in the text. The black open circles represent the observed intensities, and the calculated patterns according to the refinements are shown as red solid lines. The differences between the observed and calculated intensities are plotted at the bottom as blue solid lines. The olive, navy and red vertical bars indicate the nuclear Bragg reflections from SNWO, magnetic reflections from SNWO and NiO impurity (nuclear and magnetic reflections), respectively. (e) shows the low-$Q$ NPD patterns collected using $\lambda$ = 2.45 \AA$ $ at 1.5 K and 100 K. The refinements to the 1.5 K pattern with three different magnetic structures, by completely relaxing the direction of Ni$^{2+}$ moments or fixing it along $c$ or $b$ axes, are plotted as red solid line, blue dotted line and orange solid line(in the inset), respectively, for comparison. NPD pattern collected at 100K is normalized and plotted in green line while the 1.5K pattern is plotted as black open circles.}
	\end{figure*}

	Low-temperature NPD measurements were carried out to determine the magnetic structure of SNWO below the AFM transition at $T\rm_{N}$. Fig.~\ref{fig:NPD} shows the NPD patterns collected at 100 K (a, b) and 1.5 K (c, d) using the wavelength of 1.1545\ \AA $ $ and 2.45\ \AA, and the results of the Rietveld refinements accordingly. At 100 K, well above $T\rm_{N}$, the diffraction patterns can be perfectly fitted by the same tetragonal phase of SNWO as the room temperature (see Fig.~\ref{fig:NPD}(a, b)). Upon cooling, as shown in Fig.~\ref{fig:NPD}(c, d) for the base temperature of 1.5 K, additional reflections arising from the AFM ordering of Ni$^{2+}$ moments marked by the navy vertical bars emerge, which can be well indexed with a magnetic propagation vector $k$ = (0.5, 0, 0.5).

	To deduce the ground-state magnetic structure of SNWO, an irreducible representation analysis was performed first using the BASIREPS program integrated into the FULLPROF suite. For Ni$^{2+}$ ions locating at the 2$a$ Wyckoff position of the tetragonal lattice with the $I$4/$m$ space group, only one irreducible representation (IR) $\Gamma_1$ is possible for $k$ = (0.5, 0, 0.5) whose basis vectors are listed in Table \ref{Tab:Irep}. The three basis vectors ($\psi_{\nu}$, real unit vectors along three crystallographic axes) suggest the possibility of pointing to any direction for the Ni$^{2+}$ moments.

	\begin{table}
		\caption{\label{Tab:Irep} Basis vectors ($\psi_{\nu}$) of $\Gamma_1$, the only possible magnetic irreducible representation, for the Ni atoms occupying the 2a sites in SNWO with the space group $I4/m$ and $k$ = (0.5, 0, 0.5), obtained from representation analysis.}
		\begin{ruledtabular}
			\begin{tabular}{cccc}
			IR & $\psi_{\nu}$ & Components & Ni \tabularnewline
			\hline 
			$\Gamma_{1}$ & $\psi_{1}$ & Real & (1,0,0) \tabularnewline
			$   $ &$\psi_{2}$ & Real & (0,1,0) \tabularnewline
			& $\psi_{3}$ & Real & (0,0,1) \tabularnewline
		\end{tabular}
		\end{ruledtabular}
	\end{table}
	
	With the combination of the magnetic basis vectors, the additional magnetic reflections emerging below $T\rm_{N}$ can be fitted to determine the size and direction of the Ni$^{2+}$ moments. By simultaneous refinements to the 1.1545\ \AA $ $ and 2.45\ \AA $ $ datasets shown in Fig.~\ref{fig:NPD}(c, d), both the nuclear and magnetic structure of SNWO at the base temperature of 1.5 K can be determined with satisfactory $R$ factors. The crystallographic parameters at 1.5 K from the nuclear structure refinement is also listed in Table \ref{Table:atom info}, to compare with those at 300 K. At the base temperature, SNWO shows an overall similar tetragonal structure but an even larger octahedral rotation with the in-plane Ni-O(1)-W bond angle of 164.81$^{\circ}$, compared with 167.55$^{\circ}$ at 300 K.  
	
	According to the refined coefficients of the three basis vectors of the IR $\Gamma_1$, the Ni$^{2+}$ magnetic moment in SNWO is determined to be 1.9(2) $\mu\rm_{B}$ in size at 1.5 K, with components of $-$0.65(16), 0.69(29) and 1.70(11) $\mu\rm_{B}$ along the crystallographic $a$, $b$ and $c$ axes, respectively. Such a moment value is well consistent with that expected for the spin-only $S$ = 1 Ni$^{2+}$ moments showing an ordered moment value of $gS$ = 2 $\mu\rm_{B}$, adopting the Land\'{e} factor $g$ = 2 for spin-only cases. It is also in good agreement with the observed moment value of 1.8-2.2 $\mu\rm_{B}$ in various nickel compounds with NiO$_6$ octahedra as determined by neutron diffraction \cite{reynaud2014magnetic,Ranjith_16,su2022zigzag}, further supporting the significant quenching of the orbital moment for the Ni$^{2+}$ ions in SNWO, as evidenced by the dc magnetic susceptibility data presented in Section III. B. 

	In addition, we note that in a previous $ab$ $initio$ investigation of the magnetic ordering in SNWO, it was proposed that the AFM-II configuration with the moments aligned along the $c$ axis is energetically favorable \cite{rezaei2019ab}. In Fig.~\ref{fig:NPD}(e), we have shown the refinements to the 2.45 \AA$ $ NPD pattern collected at 1.5 K using two different magnetic structures, by completely relaxing the direction of Ni$^{2+}$ moments or fixing it along the $c$ axis, for comparison. It is clear that the latter yields a much worse agreement with the observed intensities in the low-$Q$ region ($R\rm_{mag}$ = 23.4), compared with the former ($R\rm_{mag}$ = 16.2).
	Besides, for comparison, the inset of Figure \ref{fig:NPD}(e) also displays the refinement result fixing moments along $b$ axis ($R\rm_{mag}$ = 20.2). The failure of fitting the 1.33 \AA$^{-1}$ peak also verifys the necessity of $a/c$ axis component,
	suggesting the validity of the refined result of $\vec{m}\rm_{exp}$ = ($-$0.65, 0.69, 1.70) $\mu\rm_{B}$ with components along all three crystallographic axes. 
	
\begin{figure}
		\includegraphics[width=8.6cm]{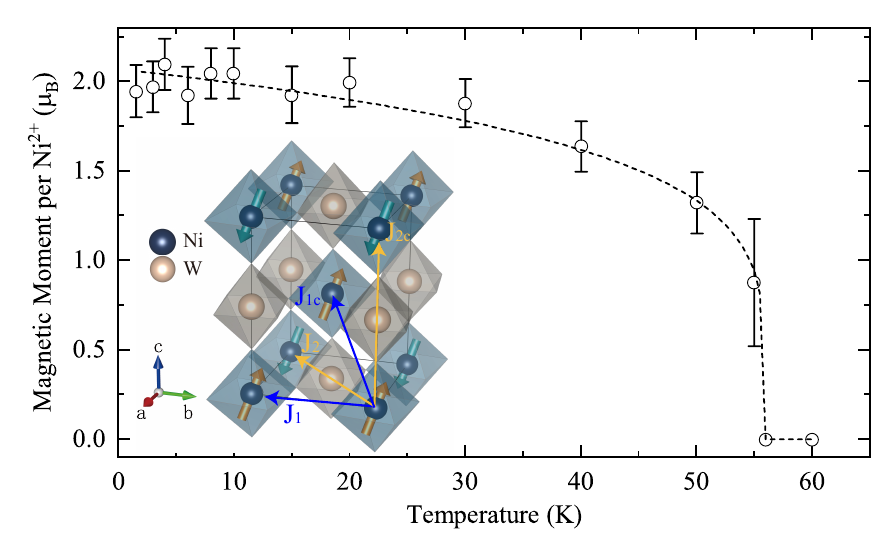}
		\caption{\label{fig:order parameter} Temperature dependence of the refined moment size of the Ni$^{2+}$ ions, where the dashed line represents a power-law fitting. The inset illustrates the experimentally determined type-II AFM structure of SNWO in the ground state, in which the blue arrows represent the in-plane and out-of-plane NN exchange couplings ($J_1$ and $J_{1c}$) and the yellow arrows represent the in-plane and out-of-plane NNN exchange couplings ($J_2$ and $J_{2c}$).}
	\end{figure}

   Fig. \ref{fig:order parameter} shows the refined moment size in SNWO as a function of temperature, which can be regarded as the order parameter associated with the AFM transition. As the dashed line presents, below $T\rm_N$ = 56 K (as determined in Section III. B), the AFM order parameter follows a power-law behavior, {$M$ $\propto$ $(\frac{T\rm_{N}-\mathit{T}}{T\rm_{N}})^{\beta}$}, with the exponent $\beta$ fitted to be 0.20(4). It is worth pointing out that the $\beta$ value determined here can not be directly compared with the universal critical exponent associated with the magnetic ordering, as Fig. \ref{fig:order parameter} does not contain enough data points in the critical region close to $T\rm_N$.

	Associated with $k$ = (0.5, 0, 0.5), the ground-state magnetic structure of SNWO is depicted as the inset of Fig. \ref{fig:order parameter}. The Ni$^{2+}$ spins align antiparallelly along both the $a$ and $c$ axes, but align parallelly along the $b$ axis, forming a type-II AFM structure. Such a type-II AFM structure was previously reported for the DP tungstates Sr$_2$CuWO$_6$ and (Ba/Sr)$_2$FeWO$_6$, ascribed to a dominant NNN interaction  $J_2$ in their spin Hamiltonian \cite{vasala2014magnetic,AZAD20021797}. The ordering pattern of the Ni$^{2+}$ spins is consistent with the AFM-II configuration theoretically predicted for SNWO in Ref. \onlinecite{rezaei2019ab}, but the moment direction differs a lot. The reason of such a discrepancy will be discussed in Section III. D.

	\subsection{First-principles calculations}
	
	\begin{figure*}[htb]
		\includegraphics[width=\linewidth]{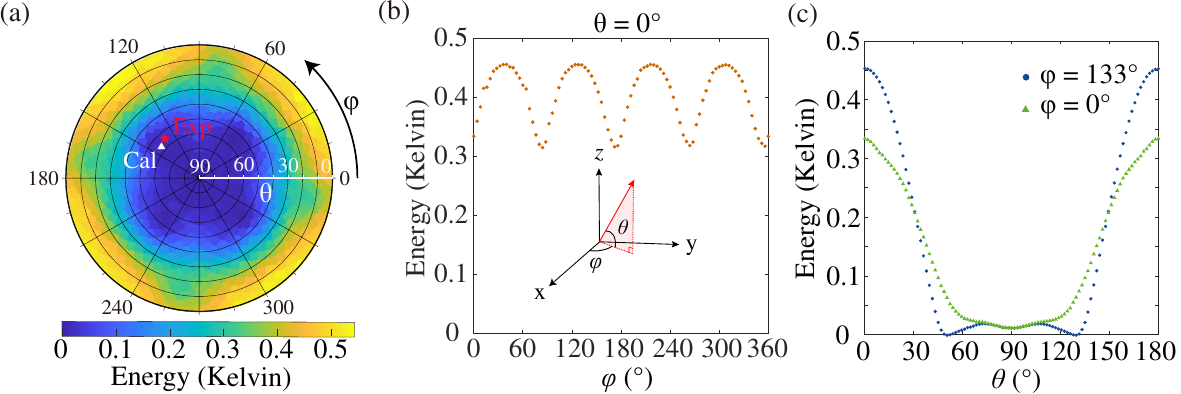}
		\caption{Angular dependence of the calculated MAE. (a) shows the 2D contour map of the MAE as functions of angles $\theta$ and $\varphi$, which are defined in the Cartesian coordinate system overlapped with the $a$, $b$, and $c$ axes of the tetragonal lattice, as shown in the inset of (b). The color represents the relative energy of certain spin orientation calculated by DFT, with respect to the magnetic ground state. The moment directions found experimentally and theoretically, according to the neutron data and the minimum of the calculated MAE, are labeled by the red star and white triangle, respectively. (b) shows the $\varphi$ dependence of the MAE (or in-plane MAE) with fixed $\theta$ = 0. (c) shows the $\theta$ dependence of the MAE (or out-of-plane MAE) with fixing $\varphi = 133^{\circ}$ (the experimentally found value) and $\varphi = 0^{\circ}$, respectively.}
		\label{fig:MAE} 
	\end{figure*}
	
		To compare with the experimentally determined magnetic structure of SNWO, the energies of 801 different spin orientations uniformly distributed in the real space have been calculated within the frame of DFT. Fig.~\ref{fig:MAE}(a) shows the 2D angular dependence of the magnetic anisotropy energy (MAE) obtained by linear interpolations to the calculated values, with respect to the magnetic ground state. By setting a Cartesian coordinate system with its $x$, $y$ and $z$ axes overlapped with the $a$, $b$, and $c$ axes of the tetragonal lattice, the direction of the Ni$^{2+}$ moments can be fully described by angles $\theta$ and $\varphi$, as illustrated in the inset of Fig.~\ref{fig:MAE}(b). The calculated petal-like MAE pattern near the center area in Fig.~\ref{fig:MAE}(a) displays a four-fold rotational symmetry around the central point ($\theta$ = 90$^{\circ}$), suggesting a $C_{4z}$ symmetry from the perspective of DFT, which is not surprising considering the tetragonal symmetry of SNWO at low temperatures. The $C_{4z}$ symmetry is better visualized in the $\varphi$ dependence of the in-plane MAE as shown in Fig.~\ref{fig:MAE}(b), with $\theta$ fixed to be 0.

		According to the calculation, the spin orientation with the moment size converged to $\vec{m}\rm_{cal}$ = ($-$0.89, 0.71, 1.65) $\mu\rm_{B}$ per formula unit owes the lowest energy, corresponding to $\theta$ = 55.5$^{\circ}$ and $\varphi$ = 141$^{\circ}$, as marked by the white triangle in Fig.~\ref{fig:MAE}(a). This calculated energy-preferred direction is highly consistent with the experimentally determined orientation of $\vec{m}\rm_{exp}$ = ($-$0.65, 0.69, 1.70) $\mu\rm_{B}$ (corresponding to $\theta$ = 60.8$^{\circ}$ and $\varphi$ = 133$^{\circ}$, marked by the red star in Fig.~\ref{fig:MAE}(a)), with a tiny difference of only 6.9$^{\circ}$. 

		As mentioned above in Section III. C, a previous DFT study proposed the AFM-II configuration with the moments aligned along the $c$ axis as the possible magnetic ground state \cite{rezaei2019ab}. However, it is clear from Fig.~\ref{fig:NPD}(e) that the experimentally determined direction of $\vec{m}\rm_{exp}$ actually deviates from the $c$ axis, with a canting angle of 29.2$^{\circ}$, which is supported by our detailed MAE calculation as shown in Fig.~\ref{fig:MAE}(a). To address the discrepancy between our work and Ref. \onlinecite{rezaei2019ab}, we have checked the $\theta$ dependencies of the calculated MAE for fixed $\varphi$ (or the out-of-plane MAE). As shown in Fig.~\ref{fig:MAE}(c), if ignoring the in-plane anisotropy and fixing the moment in the $xz$ plane ($\varphi$ = 0), the MAE curve will reach a local minimum at $\theta$ = 90$^{\circ}$, corresponding to the spin alignment along the $c$ axis, which is consistent with the result in Ref.~\cite{rezaei2019ab}. However, since the in-plane MAE shows periodic modulations (see Fig.~\ref{fig:MAE}(b)), the effect of $\varphi$ must be taken into considerations and the real minimum of the MAE has to be sought globally through the 2D mapping. By fixing $\varphi$ as the experimentally determined value of 133$^{\circ}$, the MAE curve actually reaches an even lower minimum at $\theta$ = 50.4$^{\circ}$ (see Fig.~\ref{fig:MAE}(c)), close to $\theta$ = 60.8$^{\circ}$ found experimentally, therefore supporting our model with the Ni$^{2+}$ moments canting off the $c$ axis. In addition, we note that the out-of-plane modulation of the MAE shown in Fig.~\ref{fig:MAE}(c) is much stronger, compared with that of the in-plane MAE. 

\begin{figure*}[htb]
		\includegraphics[width=\linewidth]{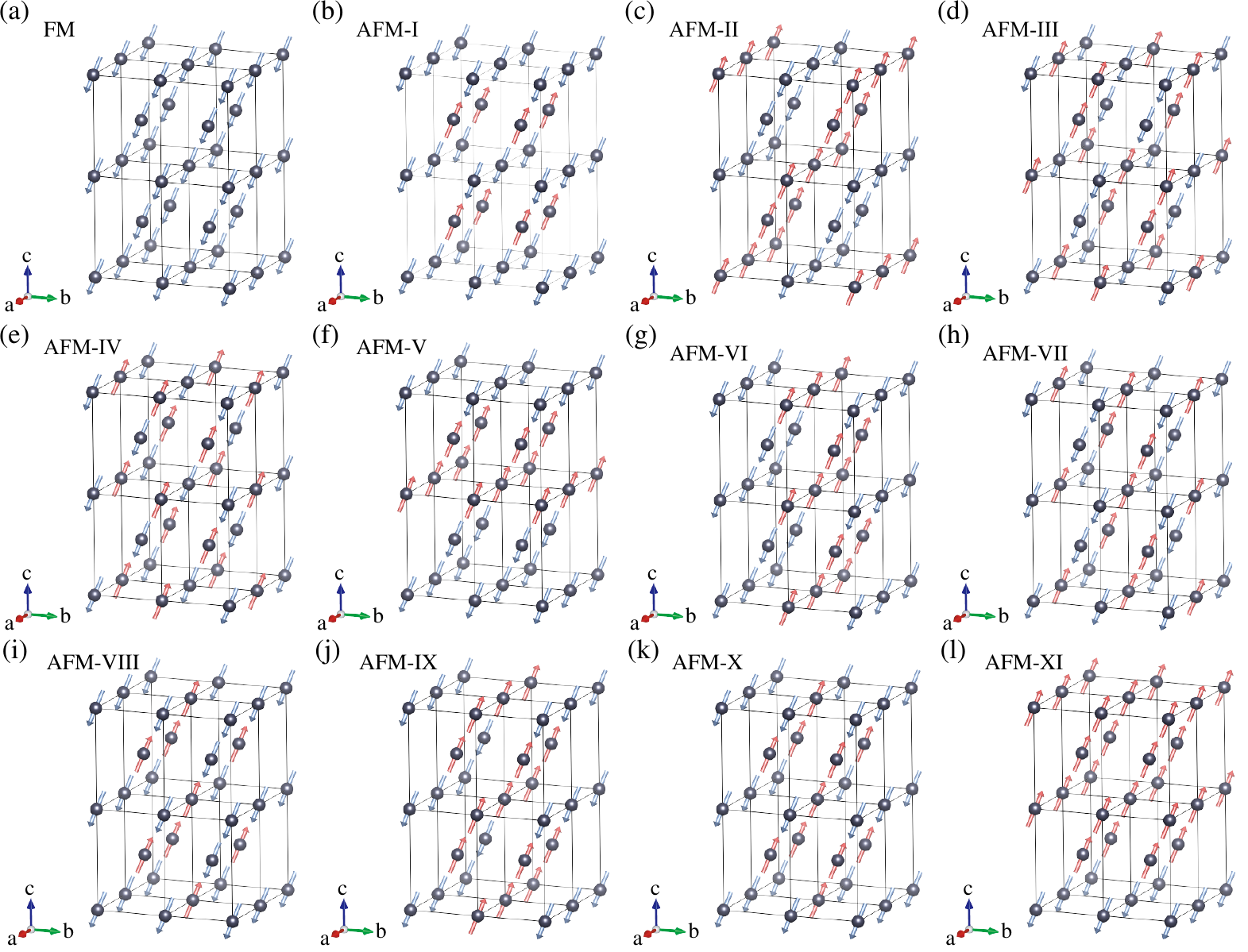}
		\caption{Twelve different representative spin configurations of SNWO, including the FM (a) and AFM-I$\sim$XI (b-l) structures.}
		\label{fig:MagConf} 
	\end{figure*}

		Furthermore, to verify the stability of the experimentally determined AFM-II structure against other possibilities, the free energies of SNWO with twelve different representative spin configurations of the Ni$^{2+}$ moments as shown in Fig.~\ref{fig:MagConf}(a-l) are calculated. The energies of these configurations with the moments all aligned along the $c$ axis have already been calculated by the GGA+$U$ method in Ref.~\cite{rezaei2019ab}. Here we have further incorporated the effect of spin-orbit coupling (SOC) and set the initial moment direction the same as $\vec{m}\rm_{exp}$, the experimentally determined one. Table \ref{tab:magconf} lists the calculation results of the relative energies, with respect to the ground state. It turns out that the lowest energy is indeed achieved for the AFM-II configuration as shown in Fig.~\ref{fig:MagConf}(c), consistent with the experimentally determined magnetic structure, further corroborating the validity of our magnetic structure model.	
 
\begin{table}[htbp]
	\centering
	\caption{Relative energies (in unit of meV) of twelve different spin configurations for SNWO, as shown in Fig.~\ref{fig:MagConf}, calculated using GGA+SOC+$U$.}
	\resizebox{\columnwidth }{!}{
	  \begin{tabular}{ccccc}
	  \hline
	  \hline
	  Spin configuration & Energy (meV) & Spin configuration & Energy (meV) \\
	 	  \hline
      FM     &    14.57  & AFM-VI  &    4.78  \\
	  AFM-I  &    13.61 & AFM-VII  &    8.89  \\ 
	  AFM-II  &    0.00  & AFM-VIII  &    7.75  \\ 
	  AFM-III  &    8.66  & AFM-IX  &    6.77  \\ 
	  AFM-IV  &    13.57 & AFM-X &    11.33 \\ 
	  AFM-V  &    9.32  & AFM-XI &    11.49 \\ 
	  \hline
	  \hline
	  \end{tabular}%
	}
	\label{tab:magconf}%
  \end{table}%

As mentioned in Section I, 	direct exchange couplings between the Ni$^{2+}$ ions are negligible because of the large distance between them, and the NN and NNN superexchange couplings occurring via a 90$^{\circ}$ Ni-O-(W)-O-Ni path and a 180$^{\circ}$ Ni-O-W-O-Ni path, respectively, have to be responsible for the magnetic ordering of the Ni$^{2+}$ moments. Based on the calculated energies of the twelve spin configurations given in table \ref{tab:magconf}, we have estimated the strengths of the superexchange couplings in SNWO, as marked in the inset of Fig. \ref{fig:order parameter}, using the mapping method similar to that used in Ref.~\cite{rezaei2019ab}. As a result, the in-plane superexchange coupling constants are estimated to be $J\rm_1 \sim$ $-$0.17 meV and $J\rm_2 \sim$ $-$2.00 meV, while the out-plane coupling constants are $J\rm_{1c} \sim$ $-$0.16 meV and $J\rm_{2c} \sim$ $-$2.46 meV. 
By a mean-field approximation, the Currie-Weiss temperature can be estimated to be
$$\theta\rm_{MF} =\frac{\it S(S+\rm1)}{3\it k\rm_B} (4\it J\rm_1 + 4\it J\rm_2 + 8\it J\rm_{1c} +2\it J\rm_{2c} ) = -115\ K, $$
based on the superexchange coupling constants estimated by our DFT calculations. This value of $\rm\theta_{MF}$ is close to $\theta = -92.8(3)\ \rm K$ extracted from the Currie-Weiss fitting as presented in Section III.B, further validating the correctness of the magnitude of our estimated coupling constants.
Furthermore, the magnitudes of the coupling constants estimated here also agree well with those calculated in Ref.~\cite{rezaei2019ab} and determined from inelastic neutron scattering in Ref.~\cite{Todate99}, indicating the dominant role of the NNN coupling $J\rm_2$ and $J\rm_{2c}$ in the spin Hamiltonian of SNWO, which is favorable for the antiparallel couplings between the Ni$^{2+}$ spins along the (001) and (110) directions and the stabilization of a type-II AFM structure. 

	Combining the results of our neutron diffraction measurements and DFT calculations, we can reach a conclusion that SNWO is a type-II antiferromagnet with a prominent difference in the magnetic anisotropy energy.  Considering the negligible spin-orbit coupling due to the significant quenching of the Ni$^{2+}$ orbital moment, this strong magnetic anisotropy is likely to arise from the single-ion anisotropy associated with the $S$ = 1 Ni$^{2+}$ ions \cite{Ramirez94, Zvyagin07, Curley21}. Another scenario responsible for the strong magnetic anisotropy might be some underlying magnetoelastic coupling \cite{Dutta21, Lee2018, Serrate07}. Further temperature-dependent XRD and NPD studies will be crucial to figure out the delicate structural change of SNWO across $T\rm_N$, including the lattice constants, the atomic displacement parameters, the NiO$_6$ octahedral rotation or distortion, and so on.

\section{Conclusion}

	In summary, we have conducted comprehensive investigations on the magnetic ordering in the double perovskite compound Sr$_2$NiWO$_6$, combining macroscopic magnetic characterizations, neutron powder diffraction measurements, and DFT calculations. Below $T\rm_N$ = 56 K, SNWO is revealed to order magnetically in a type-II AFM structure with $k$ = (0.5, 0, 0.5). Due to a significant quenching of the orbital moment, the low-temperature magnetic properties of the Ni$^{2+}$ ions can be well described by $S$ = 1 in the spin-only case, and the ordered moment at 1.5 K is determined to be 1.9(2) $\mu\rm_{B}$. In the ground state, the Ni$^{2+}$ moments in SNWO are estimated to cant off the $c$ axis by 29.2$^{\circ}$, which is well supported by the DFT calculations. In addition, the strengths of the in-plane and out-of-plane NNN superexchange couplings $J\rm_2$ and $J\rm_{2c}$ deduced from the DFT results are found to be dominant in the spin Hamiltonian of SNWO, which accounts for the stabilization of the type-II AFM structure as its magnetic ground state. 

\begin{acknowledgments}
This work is partly based on experiments performed at the Swiss Spallation Neutron Source SINQ, Paul Scherrer Institute, Villigen, Switzerland. The authors acknowledge the computational support from HPC of the Beihang University. The work at Beihang University is financially supported by the National Natural Science Foundation of China (Grant No. 12074023) and the Fundamental Research Funds for the Central Universities in China. The work at the University of Macau was supported by the Science and Technology Development Fund, Macao SAR (File Nos. 0051/2019/AFJ, 0090/2021/A2, and 0049/2021/AGJ) and the University of Macau (MYRG2020-00278-IAPME and EF030/IAPME-LHF/2021/GDSTIC). \\

\end{acknowledgments}
\bibliography{SNWO.bib}

\end{document}